\documentclass[a4paper,12pt]{article}
\usepackage{amssymb}
\usepackage{amsmath}
\usepackage{epsfig}
\usepackage{subfigure}
\usepackage{graphics}

\usepackage{latexsym}
\usepackage{rotating}
\usepackage{titlesec}

\title{The Method of ${\cal M}_{n}$-Extension: The KdV Equation}

\author{Metin G\"{u}rses \thanks{gurses@fen.bilkent.edu.tr}\\
{\small Department of Mathematics, Faculty of Science}\\
{\small Bilkent University, 06800 Ankara - Turkey}\\
Asl{\i} Pekcan \thanks{aslipekcan@hacettepe.edu.tr} \\
{\small Department of Mathematics, Faculty of Science} \\
{\small Hacettepe University, 06800 Ankara - Turkey}
}

\date{\nonumber}
\begin{document}
\maketitle
\date{\nonumber}
\newtheorem{thm}{Theorem}[section]
\newtheorem{Le}{Lemma}[section]
\newtheorem{defi}{Definition}[section]
\newtheorem{ex}{Example}[section]
\newtheorem{pro}{Proposition}[section]
\newtheorem{cor}{Corollary}[section]
\baselineskip 17pt

\numberwithin{equation}{section}

\begin{abstract}
In this work we generalize ${\cal M}_{2}$-extension that has been introduced recently. For illustration we use the KdV equation. We present five different
${\cal M}_{3}$-extensions of the KdV equation and their recursion operators. We give a compact form of ${\cal M}_{n}$-extension of the KdV equation and recursion operator of the coupled KdV system. The method of ${\cal M}_{n}$-extension can be applied to any integrable scalar equation to obtain integrable multi-field system of equations. We also present unshifted and shifted nonlocal reductions of an example of ${\cal M}_{3}$-extension of KdV.
\end{abstract}


\section{Introduction}

Obtaining new integrable systems is a very important topic in nonlinear science due to their rich structure. One of the methods to get an integrable system is using the Lax representations in algebras of higher rank. Another method is using perturbation technique preserving integrability \cite{Ma1}. In \cite{gur-pek1} we introduced a new method that we call ${\cal M}_{2}$-extension. Particularly, we considered the extensions of fifth order integrable Sawada-Kotera (SK) and Kaup-Kupershmidt (KK) equations. This method can be used to any integrable scalar equation to obtain integrable systems. Besides obtaining an integrable system we can also derive Hirota bilinear form and recursion operator of the obtained system by extending Hirota bilinear form and recursion operator of the scalar equation.

In this work we generalize ${\cal M}_{2}$-extension \cite{gur-pek1} to ${\cal M}_{n}$-extension. We use ${\cal M}_{n}$-extension method on integrable scalar equations to obtain systems of integrable equations and new integrable nonlocal equations. As an illustration, here we consider Korteweg-de Vries (KdV) equation  \cite{KdV}. Our work on the ${\cal M}_{n}$-extension of modified KdV equation is also in progress \cite{gur-pek2}. Note that multi-field extension of the KdV equation had also been studied in \cite{Athorne}-\cite{gur-kar2}.

 The ${\cal M}_{n}$-extension method consists of three main steps. The first step is to replace the dynamical variable of the integrable scalar equation by $u \to u^{0}\,\Sigma_{0} +u^{1} \Sigma_{1}+u^{2}\, \Sigma_2+\cdots+u^{n-1}\, \Sigma_{n-1}$ where $u^0=u$, $u^{i}, (i=1,2 \cdots, n-1)$ are the dynamical variables of the system. Here, the basis $\Sigma_{i}$, $(i=0, 1, 2, \cdots, n-1)$ of the ${\cal M}_{n}$ algebra satisfies the following multiplication rule:
\begin{equation}
\Sigma_{j} \cdot \Sigma_{i}=\Sigma_{i} \cdot \Sigma_{j}=f^{k}_{ij}\, \Sigma_{k},
\end{equation}
where we use summation convention for the repeated indices and $f^{k}_{ij}$ are the structural constants of the algebra which are symmetrical with respect to the indices $i$ and $j$, i.e., $f^{k}_{ij}=f^{k}_{ji}$. We obtain a system of equations for the dynamical variables $u^{i}$,  its  recursion operator and Lax pair. It is natural that these operators contain the structural constants $f^{k}_{ij}$ of the algebra.
  Second step is to obtain the symmetrical version of the system by defining new dynamical variables $q_{1}=u+u^{1}+u^{2}+\cdots+u^{n-1}$, $q_{2}=u-u^{1}+u^{2}+\cdots+u^{n-1}$, $\cdots$, $q_{n}=u+u^{1}+u^{2}+\cdots-u^{n-1}$. At the same time one can obtain the recursion operator with respect to the dynamical variables $q_{i}$. The third step is to apply consistent reductions to obtain standard (unshifted) nonlocal and shifted nonlocal reductions of the systems for $q_{i}$ \cite{gur-pek2}, \cite{abl1}-\cite{WT}. All these equations are new and integrable. Using the reduction formulas, one can obtain the recursion operators of the nonlocal differential equations. Soliton solutions of the standard nonlocal and shifted nonlocal equations can be easily obtained by using soliton solutions of the systems and reduction formulas.

\section{${\cal M}_{2}$-extension of the KdV equation}

Let $u \to U=uI+v \Sigma$ where $\Sigma^2=\alpha I+ \beta\, \Sigma$ for $\alpha=-\mathrm{det} (\Sigma)$ and $\beta=\mathrm{tr}(\Sigma)$. Hence using
\begin{equation}
U_{t}=U_{xxx}+6 U U_{x}, \label{denk0}
\end{equation}
the ${\cal M}_{2}$-extension of the KdV equation gives \cite{Ma4}-\cite{HuLiu}
\begin{eqnarray}
&&u_{t}=u_{xxx}+6 u u_{x}+6 \alpha v v_{x}, \label{denk1}\\
&&v_{t}=v_{xxx}+6 (uv)_{x}+6 \beta\, v v_{x}. \label{denk2}
\end{eqnarray}
The case when $\alpha=\beta=0$ gives first order perturbation equation of KdV \cite{Fuch}. Choosing
\begin{equation}
\Sigma=\left(\begin{array}{ll}
        0 & \alpha\cr
        1 & \beta \cr
        \end{array} \right)
\end{equation}
we obtain
\begin{equation}
{\cal R}=\left(\begin{array}{ll}
        R_{KdV} &\quad \quad \quad \alpha  (4 v+2 v_{x}\, D^{-1})\cr
        4 v+2 v_{x}\, D^{-1} &\quad \quad R_{KdV}+ \beta  (4 v+2 v_{x}\, D^{-1})\cr
        \end{array} \right).
\end{equation}

The Lax pair of the above system is given by
\begin{eqnarray}
&&L=I D^2+u I+v \Sigma, \label{lax11}\\
&&{\cal A}=4ID^3+ (6 u D+3 u_{x}) I+(6 vD+3 v_{x}) \Sigma .  \label{lax12}
\end{eqnarray}

\noindent
{\bf Remark 1}: The algebra in this {${\cal M}_{2}$-extension of the KdV equation is the unification of the algebras in the {${\cal M}_{2}$-extensions in \cite{gur-pek1}, \cite{gur-pek2}.

\vspace{0.5cm}
\noindent
{\bf Remark 2}: The functions $u$ and $v$ satisfying the system of equations (\ref{denk1}) and (\ref{denk2}) are both linear combinations of two different solutions $p=u+a v$ and $q=u-a v$ of  KdV equations $p_{t}=p_{xxx}+6 p p_{x}$ and $q_{t}=q_{xxx}+6 q q_{x}$ if $\alpha=1 +\beta$.

\section{${\cal M}_{3}$-extension of the KdV equation}

For application we use again the KdV equation.
Let $u \to U=u^{0}\, \Sigma_{0}+u^{1} \Sigma_{1}+u^{2} \Sigma_2$ where $\Sigma_{0}=I$ and $\Sigma_{1}$ and $\Sigma_{2}$ satisfy the following five different choices of the ${\cal M}_{3}$ algebra:

\vspace{0.5cm}
\noindent
{\bf I}. $\Sigma_{1}^2=0, ~\Sigma_{2}^2=0, ~\Sigma_{1} \cdot \Sigma_{2}=\Sigma_{2} \cdot \Sigma_{1}=0,$

\vspace{0.5cm}
\noindent
{\bf II}. $\Sigma_{1}^2=\Sigma_{2}, ~\Sigma_{2}^2=0, ~\Sigma_{1} \cdot \Sigma_{2}=\Sigma_{2} \cdot \Sigma_{1}=0,$

\vspace{0.5cm}
\noindent
{\bf III}. $\Sigma_{1}^2=\Sigma_{1}, ~\Sigma_{2}^2=0, ~\Sigma_{1} \cdot \Sigma_{2}=\Sigma_{2} \cdot \Sigma_{1}=0,$

\vspace{0.5cm}
\noindent
{\bf IV}. $\Sigma_{1}^2=\Sigma_{1}, ~\Sigma_{2}^2=\Sigma_2, ~\Sigma_{1} \cdot \Sigma_{2}=\Sigma_{2} \cdot \Sigma_{1}=0,$

\vspace{0.5cm}
\noindent
{\bf V}. $\Sigma_{1}^2=\Sigma_{1}, ~\Sigma_{2}^2=0, ~\Sigma_{1} \cdot \Sigma_{2}=\Sigma_{2} \cdot \Sigma_{1}=\Sigma_{2}.$\\

Here we can also write the matrix forms of the basis members $\Sigma_{1}$ and $\Sigma_{2}$, explicitly. For instance, we have
\begin{equation}
\Sigma_1=\left(\begin{array}{lll}
        0 & 0 &0 \cr
        0 & 0& 0 \cr
        0& 1& 0
        \end{array} \right), \quad \Sigma_2=\left(\begin{array}{lll}
        0 & 1 &0 \cr
        0 & 0& 0 \cr
        0& 0& 0
        \end{array} \right)
\end{equation}
for the Case I.

The method of ${\cal M}_{3}$-extension gives the following systems of KdV equations and their recursion operators corresponding to the above Cases I-V. We let $u^{0}=u$ , $u^{1}=v$, and $u^{2}=w$. We have

\vspace{0.5cm}
\noindent
{\bf I}. $u_{t}=u_{xxx}+6 u u_{x}$, $v_{t}=v_{xxx}+6 (uv)_{x}$, $w_{t}=w_{xxx}+6(u w)_{x}$, and

\begin{equation}
{\cal R}=\left(\begin{array}{lll}
        R_{KdV} &\quad\quad  0&\quad\quad   0\cr
        4 v+2 v_{x}\, D^{-1} &\quad  R_{KdV}&\quad\quad   0\cr
         4 w+2 w_{x}\, D^{-1}&\quad\quad  0 &\quad R_{KdV} \cr
        \end{array} \right),
\end{equation}

\vspace{0.5cm}
\noindent
{\bf II}. $u_{t}=u_{xxx}+6 u u_{x}$, $v_{t}=v_{xxx}+6 (uv)_{x}$, $w_{t}=w_{xxx}+6(u w)_{x}+6 v v_{x}$, and

\begin{equation}
{\cal R}=\left(\begin{array}{lll}
        R_{KdV} &\quad\quad   0&\quad\quad   0\cr
        4 v+2 v_{x}\, D^{-1} &\quad R_{KdV}&\quad\quad    0\cr
         4 w+2 w_{x}\, D^{-1}&\quad  4 v+2 v_{x}\, D^{-1} &\quad  R_{KdV} \cr
        \end{array} \right),
\end{equation}

\vspace{0.5cm}
\noindent
{\bf III}. $u_{t}=u_{xxx}+6 u u_{x}$, $v_{t}=v_{xxx}+6 (uv)_{x}+6 v v_{x}$, $w_{t}=w_{xxx}+6(u w)_{x}$, and

\begin{equation}
{\cal R}=\left(\begin{array}{lll}
        R_{KdV} &\quad\quad\quad  0&\quad\quad\quad   0\cr
        4 v+2 v_{x}\, D^{-1} &\quad R_{KdV}+4 v+2 v_{x}\, D^{-1}&\quad\quad\quad    0\cr
         4 w+2 w_{x}\, D^{-1}&\quad\quad\quad   0&\quad  R_{KdV} \cr
        \end{array} \right),
\end{equation}

\vspace{0.5cm}
\noindent
{\bf IV}. $u_{t}=u_{xxx}+6 u u_{x}$, $v_{t}=v_{xxx}+6 (uv)_{x}+6 v v_{x}$, $w_{t}=w_{xxx}+6(u w)_{x}+6 w w_{x}$, and

\begin{equation}
{\cal R}=\left(\begin{array}{lll}
        R_{KdV} &\quad\quad\quad  0&\quad\quad\quad   0\cr
        4 v+2 v_{x}\, D^{-1} &\quad R_{KdV}+4 v+2 v_{x}\, D^{-1}&\quad\quad\quad    0\cr
         4 w+2 w_{x}\, D^{-1}&\quad\quad\quad   0&\quad  R_{KdV}+4 w+2 w_{x}\, D^{-1} \cr
        \end{array} \right),
\end{equation}

\vspace{0.5cm}
\noindent
{\bf V}. $u_{t}=u_{xxx}+6 u u_{x}$, $v_{t}=v_{xxx}+6 (uv)_{x}+6 v v_{x}$, $w_{t}=w_{xxx}+6(u w)_{x}+6 (v w)_{x}$, and

\begin{equation}
{\cal R}=\left(\begin{array}{lll}
        R_{KdV} &\quad\quad\quad  0&\quad \quad\quad 0\cr
        4 v+2 v_{x}\, D^{-1} &\quad R_{KdV}+4 v+2 v_{x}\, D^{-1}&\quad \quad \quad  0\cr
         4 w+2 w_{x}\, D^{-1}&\quad 4 w+2 w_{x}\, D^{-1}&\quad R_{KdV}+4 v+2 v_{x}\, D^{-1} \cr
        \end{array} \right),
\end{equation}
where $R_{KdV}=D^2+4 u+2u_{x}\, D^{-1}$.

\vspace{0.5cm}
\noindent
{\bf Remark 3}: Here we took the five examples of Ma \cite{Ma-pla2024} for illustration and to correct the error in the Vth example. The algebra must be commutative otherwise the order of the nonlinear terms in the scalar equation becomes important. As an example in the KdV case, the terms $u u_{x}$ and $u_{x} u$ produce different systems and the integrability of the systems are not guaranteed.
\medskip

The Lax pair of the above systems I-V  are given by
\begin{eqnarray}
&&L=I D^2+uI+v \Sigma_{1}+w \Sigma_{2}, \label{lax13}\\
&&{\cal A}=4ID^3+(6u D+3 u_{x}) I+(6v D+3 v_{x}) \Sigma_{1}+(6w D+3w_{x}) \Sigma_2. \label{lax23}\
\end{eqnarray}

\section{${\cal M}_{n}$-extension}

Let $\Sigma_{i}$, $(i=0,1,2,\cdots, n-1)$ be a basis of a commutative algebra ${\cal M}_{n}$ satisfying the product rule
\begin{equation}\label{prod}
\Sigma_{i} \cdot \Sigma_{j}=\Sigma_{j} \cdot \Sigma_{i}=f^{k}_{ij}\, \Sigma_{k}.
\end{equation}
Here we use summation convention for the repeated indices and $f^{k}_{ij}$ are the structural constants of the algebra which are symmetrical with respect to the indices $i$
and $j$, i.e., $f^{k}_{ij}=f^{k}_{ji}$. Here we have $\Sigma_{0}=I$, $n \times n$ identity matrix. Hence the multiplication rule is as follows:
\begin{eqnarray}
&&I \cdot \Sigma_{i}=\Sigma_{i} \cdot I=\Sigma_i, ~~~i=0, 1,2, \cdots, n-1, \\
&&\Sigma_{a} \cdot \Sigma_{b}=f^{c}_{ab}\, \Sigma_{c}, ~~~a, b=1,2, \cdots, n-1.
\end{eqnarray}

The product defined above is associative, i.e.,
\begin{equation}
\left(\Sigma_{i} \cdot \Sigma_{j} \right) \cdot \Sigma_{k}=\Sigma_{i} \cdot \left(\Sigma_{j} \cdot \Sigma_{k} \right).
\end{equation}
In terms of the structural constants the associativity condition leads to
\begin{equation}\label{assoc}
f^{k}_{ij}\, f^{r}_{k \ell}=f^{k}_{\ell i}\, f^{r}_{k j},
\end{equation}
where $i,j,k,r,\ell=0,1,2,\cdots, n-1$. Hence we have the following theorem.

\begin{thm}
Using the method of ${\cal M}_{n}$-extension for the KdV equation, i.e., $u \to U= u^{k}\, \Sigma_{k}=u^{0} I+u^{a}\, \Sigma_{a}$ we obtain
the following system of equations:
\begin{equation}\label{ckdv1}
u^{i}_{t}=u^{i}_{xxx}+6 f^{i}_{jk}\, u^{j} u^{k}_{x},~~i=0,1,2, \cdots, n-1,
\end{equation}
or letting $u^{0}=u$ then
\begin{eqnarray}
&&u_{t}=u_{xxx}+6 u u_{x}+6f^{0}_{ab}\,u^{a}\, u^{b}_{x}, \label{ckdv2}\\
&&u_t^{a}=u^{a}_{xxx}+6 (u u^{a})_{x}+6 f^{a}_{bc} u^{b} u^{c}_{x}, \label{ckdv3}
\end{eqnarray}
where $a=1,2, \cdots, n-1$, $f^{0}_{00}=1, f^{0}_{0 a}=f^{0}_{a 0}=0$, $f^{a}_{0 b}=f^{a}_{b 0}=\delta^{a}_{b}$, and $f^{a}_{00}=0$.
 The recursion operator of the above system is given by
\begin{equation}
{\cal R}=R_{KdV} I+ \left(4 u^{a}\, +2 u^{a}_{x}\,D^{-1} \right)\, \Sigma_{a}.
\end{equation}
\end{thm}
\noindent The KdV systems above  (\ref{ckdv1}) or (\ref{ckdv2})-(\ref{ckdv3}) have been studied earlier in \cite{gur-kar1}, \cite{gur-kar2}.

If we wish to write the matrix representation of the algebra we first let $\left(\Sigma_{i} \right)^{j}_{k}=f^{j}_{ik}$. Such a representation is consistent with multiplication rule (\ref{prod}). Then we have
\begin{equation}
{\cal R}^{i}_{j}=\left(\begin{array}{ll}
        R_{KdV} &\quad \quad L^{0}_{a}\cr
        \left(4 u^{a}+2 u^{a}_{x}\, D^{-1} \right ) &\quad \quad {\cal R}^{a}_{b} \cr
        \end{array} \right),
\end{equation}
where $L^{i}_{j}=f^{i}_{jk}(4 u^{k}+2 u^{k}_{x} \, D^{-1})$,
\begin{equation}
{\cal R}^{a}_{b}=R_{KdV} \delta^{a}_{b}+ \left(4 u^{c}\, +2 u^{c}_{x}\,D^{-1} \right)\,f^{a}_{bc},
\end{equation}
and $R_{KdV}=D^2+4u+2u_{x} D^{-1}$. This result is in agreement with the works \cite{gur-kar1} and \cite{gur-kar2}.
We have the following corollaries of the theorem.

\begin{cor} In the examples given in the previous section (${\cal M}_{3}$-extension) $f^{0}_{ab}=0$. Hence all the examples considered can be written compactly as
\begin{align}
&u_{t}=u_{xxx}+6 u u_{x}, \label{ckdv4}\\
&u^{a}=u^{a}_{xxx}+6 (u u^{a})_{x}+6 f^{a}_{bc} u^{b} u^{c}_{x}, ~~a=1,2, \label{ckdv5}
\end{align}
with the recursion operators
\begin{equation}
{\cal R}=\left(\begin{array}{ll}
        R_{KdV} &\quad \quad 0\cr
        4 u^{a}+2 u^{a}_{x}\, D^{-1} &\quad \quad {\cal R}^{a}_{b} \cr
        \end{array} \right).
\end{equation}

\noindent Explicitly, the structural constants $f^{k}_{ij}$ of the ${\cal M}_3$ algebra are

\noindent  \textbf{I}. $f_{11}^0=f_{12}^0=f_{22}^0=f_{11}^1=f_{12}^1=f_{22}^1=f_{11}^2=f_{12}^2=f_{22}^2=0$,

\noindent  \textbf{II}. $f_{11}^0=f_{12}^0=f_{22}^0=f_{11}^1=f_{12}^1=f_{22}^1=f_{12}^2=f_{22}^2=0$, $f_{11}^2=1$,

\noindent  \textbf{III}. $f_{11}^0=f_{12}^0=f_{22}^0=f_{12}^1=f_{22}^1=f_{11}^2=f_{12}^2=f_{22}^2=0$, $f_{11}^1=1$,

\noindent  \textbf{IV}. $f_{11}^0=f_{12}^0=f_{22}^0=f_{12}^1=f_{22}^1=f_{11}^2=f_{12}^2=0$, $f_{11}^1=f_{22}^2=1$,

\noindent   \textbf{V}. $f_{11}^0=f_{12}^0=f_{22}^0=f_{12}^1=f_{22}^1=f_{11}^2=f_{22}^2=0$, $f_{11}^1=f_{12}^2=1$,

\noindent and the recursion operators for the systems I-V can be represented as
\begin{equation}
{\cal R}=\left(\begin{array}{lll}
        R_{KdV} &f_{1k}^0(4u^k+2u_x^k D^{-1}) & f_{2k}^0(4u^k+2u_x^kD^{-1})\\
        4 u^{1}+2 u^{1}_{x} D^{-1} & R_{KdV}+(4u^c+2u_x^c)f_{1c}^1& (4u^c+2u_x^c)f_{2c}^1 \\
        4 u^{2}+2 u^{2}_{x} D^{-1} & (4u^c+2u_x^c)f_{1c}^2 & R_{KdV}+(4u^c+2u_x^c)f_{2c}^2
        \end{array} \right).
\end{equation}
\end{cor}

\begin{cor} If $f^{0}_{ab} \ne 0$ for $n=3$, from the commutativity and associativity of $\Sigma_{0}=I$, $\Sigma_{1}$, and $\Sigma_{2}$ we obtain the following constraints on the structural constants:
\begin{align}
&1)\, f_{0j}^k=\delta_j^k,\\\
&2)\, f_{12}^0+f_{12}^1f_{12}^2=f_{22}^1f_{11}^2,\\
&3)\, (f_{12}^1)^2+f_{12}^2f_{22}^1=f_{22}^0+f_{22}^1f_{11}^1+f_{22}^2f_{12}^1,\\
&4)\, f_{12}^1f_{12}^0+f_{12}^2f_{22}^0=f_{22}^1f_{11}^0+f_{22}^2f_{12}^0,\\
&5)\, f_{12}^1f_{11}^0+f_{12}^2f_{12}^0=f_{11}^1f_{12}^0+f_{11}^2f_{22}^0,\\
&6)\, f_{12}^1f_{11}^2+(f_{12}^2)^2=f_{11}^0+f_{11}^1f_{12}^2+f_{11}^2f_{22}^2,
\end{align}
giving
\begin{align}
&f_{12}^0=-f_{12}^1f_{12}^2+f_{22}^1f_{11}^2,\label{M3-1}\\
&f_{11}^0=f_{12}^1f_{11}^2+(f_{12}^2)^2-f_{11}^1f_{12}^2-f_{11}^2f_{22}^2,\label{M3-2}\\
&f_{22}^0=(f_{12}^1)^2+f_{12}^2f_{22}^1-f_{22}^1f_{11}^1-f_{22}^2f_{12}^1.\label{M3-3}
\end{align}
Hence we have the system
\begin{align}
&u_t=u_{xxx}+6uu_x+6vv_x[f_{12}^1f_{11}^2+(f_{12}^2)^2-f_{11}^1f_{12}^2-f_{11}^2f_{22}^2]\nonumber\\
&+6(wv)_x[-f_{12}^1f_{12}^2+f_{22}^1f_{11}^2]+6ww_x[(f_{12}^1)^2+f_{12}^2f_{22}^1-f_{22}^1f_{11}^1-f_{22}^2f_{12}^1],\\
&v_t=v_{xxx}+6(uv)_x+6vv_xf_{11}^1+6(vw)_xf_{12}^1+6ww_xf_{22}^1,\\
&w_t=w_{xxx}+6(uw)_x+6vv_xf_{11}^2+6(vw)_xf_{12}^2+6ww_xf_{22}^2.
\end{align}
\end{cor}

\noindent \textbf{Example 1}: If we choose the structural constants obeying the conditions (\ref{M3-1})-(\ref{M3-3}), for instance,
$f_{22}^1=f_{11}^1=2, f_{11}^2=f_{12}^1=f_{12}^2=1$, and $f_{22}^2=-1$ giving $f_{12}^0=f_{11}^0=1$, $f_{22}^0=0$ we obtain a new KdV system as
\begin{align}
&u_t=u_{xxx}+6uu_x+6vv_x+6(wv)_x,\\
&v_t=v_{xxx}+6(uv)_x+12vv_x+6(vw)_x+12ww_x,\\
&w_t=w_{xxx}+6(uw)_x+6vv_x+6(vw)_x-6ww_x.
\end{align}

\begin{cor} To find a KdV system with four dynamical variables (${\cal M}_{4}$--extension) we let $u^{0}=u$, $u^{1}=v$, $u^{2}=w$, and $u^{3}=\rho$. Using (\ref{ckdv4}) and (\ref{ckdv5}) we get
\begin{eqnarray}
&&u_{t}=u_{xxx}+6 u u_{x}, \\
&&v_{t}=v_{xxx}+6 (uv)_{x}+6f^{1}_{bc}\, u^{b}\,u^{c}_{x}, \\
&&w_{t}=w_{xxx}+6 (uw)_{x}+6f^{2}_{bc}\, u^{b}\,u^{c}_{x}, \\
&&\rho_{t}=\rho_{xxx}+6 (u \rho)_{x}+6 f^{3}_{bc}\, u^{b}\,u^{c}_{x},
\end{eqnarray}
where $f^{a}_{bc}$ satisfy the conditions
\begin{equation}
f^{c}_{ab}\, f^{d}_{c e}=f^{c}_{ae}\, f^{d}_{cb}, ~~~a,b,c,d,e=0,1,2,3.
\end{equation}
\end{cor}
\noindent \textbf{Example 2}: A simple example is obtained by taking $f^{a}_{bc}=\delta^{a}_{3}\, s_{bc}$ where $s_{3 a}=s_{a 3}=0$. We shall consider ${\cal M}_{n}$-extension in more detail in a forthcoming publication.

\section{Nonlocal reductions}

To obtain standard (unshifted) nonlocal and shifted nonlocal reductions of the extensions of scalar integrable equations we need first to  write symmetrical form of the extensions which is the second step of the ${\cal M}_{n}$-extension method. As an example, consider the Case I of ${\cal M}_{3}$-extension of KdV equation. Let us introduce new dynamical variables $p=u+v+w$, $q=u-v+w$, $r=u+v-w$ yielding $u=\frac{1}{2}(q+r)$, $v=\frac{1}{2}(p-q)$, and $w=\frac{1}{2}(p-r)$. Let also $t\rightarrow at$, $a$ constant. Then Case I system turns to be
\begin{align}
&ap_t=p_{xxx}-\frac{3}{2}(rr_x+qq_x+(qr)_x)+3(pr+pq)_x,\label{ex-a}\\
&aq_t=q_{xxx}+\frac{3}{2}((qr)_x-rr_x+3qq_x),\label{ex-b}\\
&ar_t=r_{xxx}+\frac{3}{2}((qr)_x-qq_x+3rr_x).\label{ex-c}
\end{align}

\noindent \textbf{A.} Standard (unshifted) nonlocal reductions. Letting\\
\noindent (i) $r(x,t)=\rho_1q(\varepsilon_1x,\varepsilon_2t)$, $p(x,t)=\rho_2q(\varepsilon_1x,\varepsilon_2t)$, $\varepsilon_1^2=\varepsilon_2^2=1$, $\rho_1, \rho_2 \in \mathbb{R}$, in the three component KdV system (\ref{ex-a})-(\ref{ex-c}) gives the condition $\varepsilon_1\varepsilon_2=\rho_1=\rho_2=1$ for consistency and the
system reduces to the following nonlocal space-time reversal KdV equation
\begin{equation}\displaystyle
aq_t=q_{xxx}+\frac{9}{2}qq_x+\frac{3}{2}qq_x^{\varepsilon}+\frac{3}{2}q^{\varepsilon}q_x-\frac{3}{2}q^{\varepsilon}q_x^{\varepsilon},
\end{equation}
where $q^{\varepsilon}=q(-x,-t)$.\\

Furthermore, we have the complex unshifted reduction. Letting\\
\noindent (ii) $r(x,t)=\rho_1\bar{q}(\varepsilon_1x,\varepsilon_2t)$, $p(x,t)=\rho_2\bar{q}(\varepsilon_1x,\varepsilon_2t)$, $\varepsilon_1^2=\varepsilon_2^2=1$, $\rho_1, \rho_2 \in \mathbb{R}$, in the system (\ref{ex-a})-(\ref{ex-c}), we obtain the conditions $a=\bar{a}\varepsilon_1\varepsilon_2$, $\rho_1=\rho_2=1$ for consistency, and the
system reduces to the following nonlocal KdV equation:
\begin{equation}\displaystyle
aq_t=q_{xxx}+\frac{9}{2}qq_x+\frac{3}{2}q\bar{q}_x^{\varepsilon}+\frac{3}{2}\bar{q}^{\varepsilon}q_x-\frac{3}{2}\bar{q}^{\varepsilon}\bar{q}_x^{\varepsilon},
\end{equation}
where $\bar{q}^{\varepsilon}=\bar{q}(\varepsilon_1x,\varepsilon_2t)$. The above equation consists three different nonlocal equations; nonlocal space reversal KdV equation for $(\varepsilon_1,\varepsilon_2)=(-1,1)$ with $a=-\bar{a}$; nonlocal time reversal KdV equation for $(\varepsilon_1,\varepsilon_2)=(1,-1)$ with $a=-\bar{a}$; nonlocal space-time reversal KdV equation for $(\varepsilon_1,\varepsilon_2)=(-1,-1)$ with $a=\bar{a}$.

\medskip
\noindent \textbf{B.} Shifted nonlocal reductions. Similarly, we can introduce shifted nonlocal reductions. Letting\\
\noindent (i) $r(x,t)=\rho_1q(\varepsilon_1x+x_0,\varepsilon_2t+t_0)$, $p(x,t)=\rho_2q(\varepsilon_1x,\varepsilon_2t)$, $\varepsilon_1^2=\varepsilon_2^2=1$, $\rho_1, \rho_2, x_0, t_0 \in \mathbb{R}$, in the KdV system (\ref{ex-a})-(\ref{ex-c}), we get $\varepsilon_1\varepsilon_2=\rho_1=\rho_2=1$ for consistency. Hence the
system reduces to the shifted nonlocal space-time reversal KdV equation given by
\begin{equation}\displaystyle
aq_t=q_{xxx}+\frac{9}{2}qq_x+\frac{3}{2}qq_x^{\varepsilon}+\frac{3}{2}q^{\varepsilon}q_x-\frac{3}{2}q^{\varepsilon}q_x^{\varepsilon},
\end{equation}
where $q^{\varepsilon}=q(-x+x_0,-t+t_0)$.\\

We have also the complex shifted reduction. Letting\\
\noindent (ii) $r(x,t)=\rho_1\bar{q}(\varepsilon_1x+x_0,\varepsilon_2t+t_0)$, $p(x,t)=\rho_2\bar{q}(\varepsilon_1x+x_0,\varepsilon_2t+t_0)$, $\varepsilon_1^2=\varepsilon_2^2=1$, $\rho_1, \rho_2, x_0, t_0 \in \mathbb{R}$, in the KdV system (\ref{ex-a})-(\ref{ex-c}), we obtain the conditions $a=\bar{a}\varepsilon_1\varepsilon_2$, $\rho_1=\rho_2=1$ for consistency. Therefore, the system reduces to the following shifted nonlocal KdV equation:
\begin{equation}\displaystyle
aq_t=q_{xxx}+\frac{9}{2}qq_x+\frac{3}{2}q\bar{q}_x^{\varepsilon}+\frac{3}{2}\bar{q}^{\varepsilon}q_x-\frac{3}{2}\bar{q}^{\varepsilon}\bar{q}_x^{\varepsilon},
\end{equation}
where $\bar{q}^{\varepsilon}=\bar{q}(\varepsilon_1x+x_0,\varepsilon_2t+t_0)$. Here we have three different shifted nonlocal equations; shifted nonlocal space reversal KdV equation for $(\varepsilon_1,\varepsilon_2)=(-1,1)$ with $a=-\bar{a}$, $t_0=0$; shifted nonlocal time reversal KdV equation for $(\varepsilon_1,\varepsilon_2)=(1,-1)$ with $a=-\bar{a}$, $x_0=0$; shifted nonlocal space-time reversal KdV equation for $(\varepsilon_1,\varepsilon_2)=(-1,-1)$ with $a=\bar{a}$.

\section{Concluding remarks}
In a recent work \cite{gur-pek1} we introduced ${\cal M}_{2}$-extension which is used to obtain new integrable systems from known integrable scalar equations. In this work we generalized our work to ${\cal M}_{n}$-extension. For illustration we considered KdV equation. Obtaining such integrable systems is important since by using standard (unshifted) nonlocal and shifted nonlocal reductions we can obtain new integrable nonlocal equations. We indeed presented an example of ${\cal M}_{3}$-extension of KdV equation, its unshifted nonlocal and shifted nonlocal reductions. The method of ${\cal M}_{n}$-extension can be used to any scalar integrable equation to produce integrable coupled systems of integrable equations. In particular application to non-polynomial integrable scalar equations will be interesting.

\section{Acknowledgment}
  This work is partially supported by the Scientific
and Technological Research Council of Turkey (T\"{U}B\.{I}TAK).

\end{document}